# First Workshop On the Globalization of Modeling Languages (GEMOC 2013) *

Benoit Combemale[1] and Julien De Antoni[2] and Robert B. France[3] and Frédéric Boulanger[4] and Sébastien Mosser[5] and Marc Pantel[6] and Bernhard Rumpe[7] and Rick Salay[8] and Martin Schindler[9]

[1] University of Rennes 1, France
[2] University of Nice Sophia Antipolis, France
[3] Colorado State University, USA
[4] Supélec, France
[5] Polytech'Nice Sophia Antipolis, France
[6] INPT ENSEEIHT, France
[7] RWTH Aachen University, Germany
[8] University of Toronto, Canada
[9] MaibornWolff, Germany

**Abstract.** The first edition of GEMOC workshop was co-located with the MODELS 2013 conference in Miami, FL, USA. The workshop provided an open forum for sharing experiences, problems and solutions related to the challenges of using of multiple modeling languages in the development of complex software-based systems. During the workshop, concrete language composition artifacts, approaches, and mechanisms were presented and discussed, ideas and opinions exchanged, and constructive feedback provided to authors of accepted papers. A major objective was to encourage collaborations and to start building a community that focused on providing solutions that support what we refer to as the globalization of domain-specific modeling languages, that is, support coordinated use of multiple languages throughout the development of complex systems. This report summarizes the presentations and discussions that took place in the first GEMOC 2013 workshop.

## 1 Introduction

Modern software-intensive systems serve diverse stakeholder groups and thus must address a variety of stakeholder concerns. These concern spaces are often associated with specialized description languages and technologies that are based on concern-specific problem and solution concepts. Software and system engineers are thus faced with the challenging task of integrating the different languages and associated technologies used to produce various artifacts in the different concern spaces.

GEMOC 2013 was a full-day workshop that brought together researchers and practitioners in the modeling languages community to discuss the challenges associated with integrating multiple, heterogeneous modeling languages. Supporting coordinated use

---

* This workshop is supported by the GEMOC initiative (http://gemoc.org) and the ReMoDD initiative (http://www.cs.colostate.edu/remodd)

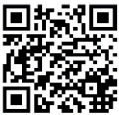

of modeling languages leads to what we call the globalization of modeling languages. The languages of interest for the participants ranged from requirements to runtime languages, and included both general-purpose and domain-specific languages. Challenges related to engineering composable languages, semantic composition of languages and to reasoning about systems described using heterogeneous languages were discussed

The workshop GEMOC 2013 was co-located with MODELS 2013 in Miami, FL, USA, on September 29th, 2013. In this report we document the various presentations, as well as the enthusiastic and intense discussions that took place during the workshop. The workshop report is organized as follows. Section 2 gives a broad overview of the workshop, including the topics of interest and relevant application domains. Section 3 describes the paper review and selection process, and the structure of the workshop. Section 4 summarizes the presentations made in the first half of the workshop day. Section 5 then presents the main points raised in the discussions that followed the presentations. The intent of the discussions was to position each presented approach in a language and model composition landscape. An initial global landscape was introduced and discussed in the second half of the workshop day. The results of that discussion are summarized in Section 6. Conclusions are drawn in Section 7.

## 2 Workshop Overview

Software intensive systems are becoming more and more complex and interconnected. Consequently, the development of such systems requires the integration of many different concerns and skills. These concerns are usually covered by different languages, with specific concepts, technologies and abstraction levels. While the use of multiple languages eases the development of target concerns, it also raises language and technology integration problems at the different stages of the software life cycle. In order to reason about the global system, it becomes necessary to explicitly describe the different kinds of the relationships that can exist between the different languages used in the development of a complex system. To support effective language integration, there is a pressing need to reify and classify these relationships, as well as the language interactions that the relationships enable.

In this context, the workshop attracted submissions that include outlines of language integration approaches and case studies, or that identify and discuss well-defined problems about the management of relationships between heterogeneous modeling languages. The goal was to facilitate discussions among the participants that lead to an initial classification of useful kinds of language relationships and their management.

The Call for Papers explicitly solicited contributions that describe case studies on coordinated use of multiple modeling languages, and/or practical experience, opinions and related approaches. Authors were invited to submit short papers describing (i) their language integration experience, or (ii) novel approaches for integrating modeling languages. Authors were also invited to store full versions of models used to illustrate their novel approach or experience in the Repository for Model Driven Development (ReMoDD). This allowed us to share the actual models with participants and the wider modeling community before, during and after the workshop.

The topics of interest for GEMOC 2013 was:

- Composability and interoperability of heterogeneous modeling languages
- Language integration challenges, from requirement to design, to analysis and simulation, to runtime.
- Model and metamodel composition
- Multi-paradigm modeling and simulation

Submissions describing practical and industrial experience related to the use of heterogeneous modeling languages were also encouraged, particularly in the following application domains:

- Cyber-Physical Systems, System of Systems
- Smart City, Smart Building, Home automation
- Complex Adaptive Systems
- Internet of Services, Internet of Things

## 3 Workshop Organization

Benoit Combemale, Julien Deantoni and Robert B. France organized and chaired the program committee (PC) for this GEMOC edition. The workshop website[10] and the call for papers (CfP) was online, quite 6 months before the workshop date. Apart from the workshop website, the CfP was announced on different professional mailing lists (e.g., SEWORLD, pUML, planetmde).

We received 11 submissions and finally accepted 8 papers, resulting in an acceptance rate of 72%. Each submission was reviewed by at least three members of the PC. Papers were selected based on their relevance to the topics for the workshop and the reviews provided by PC members. The organizers are very grateful to PC members for performing this important service to the community and for the quality of their work. The PC consisted of: Walter Cazzola, (DICo, University of Milano, Italy), Benoit Combemale (University of Rennes 1, France), Julien DeAntoni (University of Nice Sophia Antipolis, France), Robert B. France (Colorado State University, USA), Jeff Gray (University of Alabama, USA), Jean-Marc Jézéquel (University of Rennes 1, France), Jörg Kienzle (McGill University, Canada), Marjan Mernik (University of Maribor, Slovenia), Pieter J. Mosterman (MathWorks, USA), Gunter Mussbacher (University of Ottawa, Canada), Bernhard Rumpe (RWTH Aachen University, Germany) and Eugene Syriani (University of Alabama, USA).

The format of the workshop reflected the goals of the workshop: To provide constructive feedback on submitted papers and models on the coordinated use of different modeling languages, and to foster collaborations and community building. The format of the workshop was that of a working meeting. Hence, it was less focused on presentations and more on producing and documenting a research roadmap that identifies challenges, different forms of language integration, and relates existing solutions.

The workshop day was split into two parts. In the first part, an introduction and short presentations of the accepted papers were given. The second part of the day was dedicated to open discussions of the presented contributions and other related topics. We lead the discussion towards a classification of existing and proposed forms of language integration. This workshop report is the result of the beneficial discussions we had.

---

[10] Cf. http://gemoc.org/gemoc2013

## 4 Papers and Models Summary

The papers and the associated talks are available on the workshop web pages (`http://www.gemoc.org/gemoc2013`). The associated models are available in the ReMoDD repository (`http://www.remodd.org`).

*[1] Toward Denotational Semantics of Domain-Specific Modeling Languages for Automated Code Generation (Danielle Gaither and Barrett R. Bryant: University of North Texas)* This contribution advocates the use of denotational semantics instead of operational ones in order to specify, at the language level, the behavior of models defined using Domain Specific Modeling Languages. The key purpose is to ease the development of Automated Code Generators from these languages through their denotational semantics. Denotations are provided as functions that interprets the models by manipulating metamodel's elements. In future work, the use of mathematical functions should ease the composition of language semantics through function composition. This contribution is applied on a restricted subset of a dedicated modeling language for Role Playing Games.

*[2] From Sensors to Visualization Dashboards: Challenges in Languages Composition (Sebastien Mosser, Ivan Logre and Philippe Collet: University Nice-Sophia Antipolis ; Nicolas Ferry: SINTEF IKT)* This contribution describes a full fledged use case from the Internet of Things based on the SensApp platform: bikes equipped with sensors and ad-hoc networks transmit different kind of data that are gathered, analysed and forwarded to end users, which can configure their own interfaces to browse the consolidated data. The authors' purpose is to describe the use case and the associated issues, not to advocated a specific solution. The use case contains eleven kinds of models that must communicate and cooperate: Topology, Behavior, Communication, Component, Data, Computation, Resource, Requirement, Task, User Interface and Variability. All these models must be consistent and reusable: sophisticated composition operators are thus needed to build the whole system model from the concern models. The authors identify some relations between the models: co-exists with, uses, constraints, implements. An interesting question is then: are these relation between languages or models?

*[3] Heterogeneous Model Composition in ModHel'X: the Power Window Case Study (Frédéric Boulanger, Christophe Jacquet and Cécile Hardebolle, Supélec)* This contribution details the use of the ModHel'X toolset for heterogeneous model execution for a Car Power Window system. Strongly inspired by Ptolemy, ModHel'X provides Multi Paradigm Modeling focusing on the semantic adaptation between the various involved Models of Computation. They use the Tagged Event Specification Language (TESL): a DSML for expressing time and control constraints between different heterogeneous parts of the system. TESL takes the synchronous part of the Clock Constraints Specification Language (CCSL), which is part of the UML/MARTE standard, and extends it with time tags on events and relations between the time scale of clocks. The purpose of selecting the synchronous subset of CCSL is to ease the implementation of tools around TESL. Adding a time tag to events, and establishing relations between time tags is more

efficient for simulation because it allows arbitrary precision on the date of events without requiring high frequency clocks. The Car Power Window models include Discrete Event (communication between parts), Timed Finite State Machine (controller) and Synchronous Data Flow (mechanical parts) with adaptation between DE and TFSM, and between DE and SDF.

*[4] Railroad Crossing Heterogeneous Model (Matias Ezequiel Vara Larsen, University Nice-Sophia Antipolis ; Arda Goknil, INRIA Sophia Antipolis)* This contribution focuses on the composition of heterogeneous models applied to a Railroad Crossing Management System (RCMS). Models are conforming to languages expressed with four language units: Abstract Syntax (AS), Domain Specific Actions (DSA), Domain Specific Events (DSE) and Model of Concurrency (MoC). These units are put in consistency through the DSE, which are expressed using the Event Constraint Language (ECL), an extension of OCL with events and event relations. From an ECL specification on a language, an automatic transformation of a model is provided to create the CCSL constraints, i.e. the execution model of the model. These constraints are then solved to drive the execution of the models. Composition operators express coordination between the DSE of two or more languages and are used to create constraints written in CCSL that are combined with the ones provided by the execution models of the languages. The whole approach is illustrated with the Rail Crossing Management System that combines a Barrier Detection Controller modelled using Timed Finite State Machine and a Barrier Motor Controller modelled using fUML.

*[5] On the Challenges of Composing Multi-View Models (Matthias Schöttle and Jörg Kienzle: McGill University)* This contribution targets the specification of complex systems using separation of concerns. The key aspects is a strategy to integrate metamodels and the associated model composers to build multi-view formalisms including multi-view composers. Each metamodel is provided with an internal model composer. This proposal relies on an asymmetric approach: one of the metamodels is selected as independent metamodel which is unchanged, and the other metamodel is integrated in this one. The independent metamodel composer is thus unchanged, and the whole multi-view composer is derived automatically from the integration of the second metamodel. This proposal is applied to the Touch RAM toolset implementing Reusable Aspect Models that allow specifying and reusing concerns in the development of complex systems. RAM provides structural, behavioural and protocol modelling using class, sequence and state machine diagrams.

*[6] Using partial model synthesis to support model integration in large-scale software development (Marsha Chechik and Rick Salay: University of Toronto)* This contribution targets the synthesis of model stubs that enable the simulation of models specified using interface contracts. In the development of complex systems with many stakeholders, model interface contracts allow to loosen development schedule constraints. However, these contracts cannot be executed and thus the whole system can only be simulated when all models have been defined. The synthesis of models satisfying those contracts provide stubs that can be used in that purpose. These models can be refined incrementally when additional data are available during the development. This proposal

is applied to the models of a webmail system relying on sequence diagrams to describe interaction scenarios and Fluent Linear Temporal Logic to express system invariants. Other experiments are planned to extend the proposal to other kind of models of computation.

*[7] Enhance the Reusability of Models and Their Behavioral Correctness (Papa Issa Diallo and Joël Champeau: ENSTA Bretagne)* This contribution targets the preservation of model semantics during a Model Driven Engineering process that combines several executable languages and simulation tools using model transformations from languages to languages. In that purpose, the authors rely on the COMETA framework to specify high level Models of Computation relying on the low level MoC provided by each model execution toolset. This approach constrains the usual execution of the models inside a given tool in order to provide a common semantics for the various models of the same system during the development phases. This proposal is applied to the integration of two toolsets involved in the development of safety critical systems: Rhapsody UML and ForSyDe-SystemC.

*[8] Black-box Integration of Heterogeneous Modeling Languages for CPS (Markus Look, Antonio Navarro Perez, Jan Oliver Ringert, Bernhard Rumpe and Andreas Wortmann: RWTH Aachen University)* This contribution targets the integration of six independently developed modeling languages and its application to the robotics domain. These languages includes a component & connector architecture description language, automaton, I/O table, class diagrams, OCL, and a Java DSL. The resulting MontiArcAutomaton modeling framework allows to model the logical software architecture and the system behavior of robotics applications. Even the languages were not developed for the robotics domain in the first place, the existing language components could be completely reused using the language composition approaches of the MontiCore framework.

## 5 Why, What, Where, How Composing Languages and Models?

An important part of the workshop was dedicated to discussions on how to realize the vision of globalized modeling languages. The aim was to map out a landscape of language and model composition issues that reflected ideas raised in the previous presentations and the experiences of workshop participants. To structure the discussions, the organizers proposed to follow the Why? What? Where? How? pattern as reported in the rest of this section.

### 5.1 Why?

Composition of models is highly desired, because the history of computer science has shown that development can only be managed by dividing a complex task or product into smaller easier sub-structures. This in particular holds for models which we like to decompose along sub-structures of the product, e.g. into components [2], or along viewpoints that allow us to look at the same components from different perspectives [5]. However, decomposition is useful only if the different realizations obtained through the design of the subsystems, can be composed again to obtain a realization of the whole system. Structural decomposition of models needs a composition operator present within the modeling language, for example, in finite state machines, where composing two state machines leads to another one or in class diagrams, where diagrams are basically merged along classes with the same name.

When different viewpoints are modeled in different languages, model composition also becomes language composition, where models of different languages need to become interoperable. This is necessary for example for the UML, where structural and behavioral models describe the same system, but also in the recently emerging and potentially more interesting domain of distributed systems, where several different, but similar languages are used to describe the behavior of system components [2].

We can identify four main reasons for using models written using different formalisms:

- The decomposition of a complex system into different parts that belong to different technical domains which have their own modeling formalisms and tools;
- The change in the level of abstraction during the design of a system, for instance from a non-deterministic automata for specification, to VHDL for the concrete realization;
- The need for different models of a system that model different aspects, or offer differents views on the system and therefore use different modeling formalisms: the timing or power consumption model of a system may use a different formalism than the functional model;
- The need for different models of a system for different activities during the design process: a model used for code generation and a model used for verification may use different formalisms.

These four reasons lead to four corresponding issues:

- How does one compose the structure and behavior of heterogeneous models?
- How does one check that a model is a refinement of a more abstract model?
- How does one synchronize different views on different aspects of a system?
- How does one check the consistency of different models of a given system?

Only if these forms of composition are well understood will we be able to carry out integrated code generations, simulations, validations and verifications on integrated models. These composition issues are rooted in the semantics of the models, with the added complexity of the definition of a mapping between concepts that belong to different semantic domains and are represented according to different syntaxes.

Even though the composition of models and languages is somehow related to the forms of composition for products, we need to be explicitly aware that these are different forms of compositions. That is why it is necessary to explicitly clarify what the composition is about. In particular, it would be very nice if a specification formalism provides a composition operator for its specifications in such a way that this composition conforms to the composition of the specified components.

Identifying the relationships that exists between different languages is useful to support the understanding of language composition, from a systemic point of view [2]. Understanding this kind of relations leads to a classification of existing composition relationships, e.g., uses (a computation model uses a data model), implements (a task model implements a use case). An off-the-shelf classification of composition in the context of DSML provides guidance to engineers who have to perform such compositions.

Additionally, besides the technical forms of composition, the ability to compose has also an organizational consequence. If we can compose individually developed models, we can decompose the team into smaller sub-teams developing individual parts. This is why composition of models is an important requisite for successful projects developing complex products.

### 5.2 What?

Formally, composition is an operator taking at least two arguments and producing one result. Model composition thus involves manipulation of models. Each model has a syntax and a semantics. Composition can operate on the syntax, for example deriving an integrated new model describing the information originally contained in both models [8]. Composition could also work on the semantics only [4]. This, for example, works well with denotational semantics, when the models are mapped onto a common semantic domain and the composition is denoted using composition in the semantic domain only [1]. As another possibility we could explain composition at the code generation level, by explaining how the code, which has been derived from each model individually, is linked together [7]. In any case, this usually requires an understanding of the interfaces between models, and respectively, their derivatives [4,3,7].

Furthermore, since models are developed incrementally across the development lifecycle and at different rates in different sub-teams, requiring models to be complete before composition can cause project delays. To address this we must also allow for the meaningful composition of partially complete models [6]. Such compositions must preserve the information that is known without biasing the information that is still unknown.

Relying on an existing classification of existing composition supports the identification of the elements to compose during the process of designing a given composition. One can rely on the existing relations to guide its own development and support incremental approaches. The fact that a language L "implements" concepts modeled by a language L' implies that both elements (the concept to be implemented in L' and its associated artefacts in L) exists at the time of the composition, even if developed asynchronously. On the contrary, if L "uses" elements from L', one cannot use L to work while the model in L' is not complete.

These considerations hold for any kind of modeling languages, structural or behavioral. However, from a practical point of view, in a top-down approach we develop structure first and thus need decomposition on structure on the higher level. Typically behavioral aspects are only modeled and thus composed on a more fine-grained details level in the later phases of the development. On the other hand, a bottom-up approach involves composing already developed parts of a system to produce a complete system. However, in bottom-up approaches, the composition is usually made at the model level by making interoperable homogeneous interfaces and usually do not address composition of syntactic elements.

### 5.3 Where?

We outlined in the previous section how the composition can operate on different parts of languages and models. It is also interesting to note that the composition can be specified and applied at different level: From the language level to some compiled code. Additionally, the specification of the composition and its application can also be done at different levels. For instance, a composition operator can be specified between two languages in order to create a new language [8]. In this case, both the specification and the application of the composition is done at the language level. Another composition operator could be specified between two languages in order to create relations between the models of the respective languages, but without explicitly creating a new language [4]. It is important to identify at which level the specification and the application of the composition is made to understand the impact on the reuse of the composed language tooling. On this point, several aspects of the involved languages have to be considered:

- concrete and abstract syntax
- semantics
- consistency checks
- code generators

This also includes their resulting infrastructure, for example, lexer, parser, symbol tables, or editors. For an efficient and agile composition of existing languages, these aspects should be reused as much as possible and only a minimal amount of additional glue-code should be necessary. The level at which the specification and the application of the composition is realized depends on the main objective of the composition (i.e. the why). For instance, because the semantics of a language combination has to be considered, the composition is preferably defined on the language level. However, in a reuse perspective the linkage should take place as late as possible to be able to provide the same language component for different compositions. The idea behind this is the same as for framework or libraries in programming languages which can be used as pre-compiled components without providing the sources. This speeds up the development process significantly as the libraries do not have to be compiled again and again for every usage while it restrains the possibility of analysis of the composed system.

### 5.4 How?

In order to be able to compose different languages and models without developing the languages for each combination from scratch again or specifying the matching parts for each model combination we need a concise and consistent definition of the interfaces of languages and models. An approach similar to that used in existing programming languages, where methods or classes have an interface for its usage encapsulating the complexity of the actual implementation, may be applicable. If languages provide interfaces that hide their internal complexity, a library of language components could be created allowing the composition of languages by writing a minimal amount of glue-code for each combination.

The need for an interface holds for models as well. Each model should provide an interface which can be used from other models without knowing the internal details [3,6,7]. This means that the modeling languages have to provide two concepts: one for describing the interface of a model and one for referencing or calling such interfaces. These concepts can even be used to combine models whose languages were originally developed independently. To give an example let us take the concept of a method call in a programming language like Java which was originally designed as a reference to a method signature within the same language. However, this concept could be reused to call other interfaces as well, e.g. an interface to a Statechart. This only requires that a method call allows one to specify all aspects requested by the Statechart interface. If this holds only a mapping of the concepts of the reference (the method call) and the targeted interface (of the Statechart) is needed. In this way two languages can be composed by adding an additional semantics to an existing language concept without changing the original language itself. It can also be seen as scheduling the new events from the statechart (start, entry in a state, trigger, etc) together with the events of the original java program and specifically here with the events of its method call (call, return)[4,5].

By using such approaches, a composition of languages can be reduced to a mapping between the language concepts needed to combine models along their interfaces. Only if the existing language concepts are not sufficient to define such a mapping, an extension of one or both languages is needed for the composition.

And finally the composition of the language tooling have to be considered. However, if all of these aspects are implemented using clear interfaces, the composition could be derived from the already defined mapping of the language concepts even automatically.

When only partial information is known about a model, techniques such as model synthesis can be used to construct an approximate, but well-formed, model based on the known information. Such approximate models can be used as stand-ins for partially complete models in composition operations [6].

The various criteria about compositions lead to various implementations. Such implementations can be hard coded [6,8], based on predefined operators [5], or specified thanks to a dedicated DSL [4,3] Additionally, many important properties can be used to characterise the implementation, for example implementations may be asymmetric or symmetric, or may be transient or not. These properties should be clarified and we should understand how the why of the composition is influencing such choices.

## 6   Conclusion

The first edition of the workshop met its goals and thus can be considered a success. The discussions that took place during the workshop were of very high quality and provided insights into some of the pressing problems associated with the use multiple modeling languages in development projects. A key result is a deep appreciation by the participants of the need to develop support for globalizing modeling languages. The ongoing research that was reported in the workshop and the discussions that took place are a good indication that community around these problems is emerging. For more information about the GEMOC initiative please visit the following website: http://gemoc.org.

## 7 Acknowledgments

GEMOC 2013 was supported by the GEMOC initiative (cf. http://gemoc.org) that promotes research seeking to develop the necessary breakthroughs in software languages to support global software engineering, i.e., breakthroughs that lead to effective technologies supporting different forms of language integration, including language collaboration, interoperability and composability. Part of this initiative, the workshop was partially supported by the ANR INS project GEMOC (grant #ANR-12-INSE-0011).

GEMOC 2013 was also supported by the ReMoDD project, an NSF funded project (award number CNS-1305381).